# TOWARDS THE NEURAL CODE ENDOWED IN CORTICAL MICROCOLUMNS

Arturo Tozzi
Center for Nonlinear Science, University of North Texas
1155 Union Circle, #311427
Denton, TX 76203-5017, USA, and
Computational Intelligence Laboratory, University of Manitoba, Winnipeg, Canada
Winnipeg R3T 5V6 Manitoba
ASL NA2 Nord
tozziarturo@libero.it

James F. Peters
Department of Electrical and Computer Engineering, University of Manitoba
75A Chancellor's Circle, Winnipeg, MB R3T 5V6, Canada and
Department of Mathematics, Adıyaman University, 02040 Adıyaman, Turkey,
Department of Mathematics, Faculty of Arts and Sciences, Adıyaman University
02040 Adıyaman, Turkey and Computational Intelligence Laboratory, University of
Manitoba, WPG, MB, R3T 5V6, Canada
james.peters3@umanitoba.ca

Ottorino Ori
Actinium Chemical Research, Via Casilina 1626/A, 00133 Rome, Italy
ottorino.ori@gmail.com

**ABSTRACT**

Artificial neural systems and nervous graph theoretical analysis rely upon the stance that the neural code is endowed in logic circuits, *e.g.*, spatio-temporal sequences of ON/OFF spiking neurons. Nevertheless, this assumption does not fully explain complex brain functions. Here we show how nervous activity, other than logic circuits, could instead depend on topological transformations and symmetry constraints occurring at the micro-level of the cortical microcolumn, *i.e.*, the embryological, anatomical and functional basic unit of the brain. Tubular microcolumns can be flattened in guise of a fullerene-like two-dimensional lattices, equipped with about 80 nodes, standing for pyramidal neurons, where neural computations take place. We show how the countless possible combinations of activated neurons embedded in the lattice resemble a barcode. Different assemblies of firing neurons might stand for diverse codes, each one responsible for a single mental activity. A two-dimensional fullerene-like lattice not just simulates the real microcolumn's microcircuitry, but also allows us to build artificial networks equipped with robustness, plasticity and fastness, because they are grounded on simple topological changes corresponding to pyramidal neurons' activation.

Current computational brain models, based on neural networks performing logic operations (Izhikevich, 2010; Sporns 2013; Ursino et al., 2014), are not able to fully elucidate a large repertoire of brain functions and mental faculties, such as attention and perception, emotions and cognition, memory and learning, higher cognitive processes (decision making, goal-directed choice, etc.) (Gazzaniga, 2009), mind wandering (Andrews-Hanna et al., 2014) and so on. In order to build a versatile network able to simulate the brain function at micro-levels of observation, we introduce a cortical model borrowed from fullerene's geometry, *e.g.*, a method able to evaluate symmetry constraints and topological indices for micro-structures (Koorepazan-Moftakhar et al., 2015). Although its primary application concerns the description of carbon-networks in chemical compounds, this method provides a mathematical treatment that can be used in order to rank topological invariants in the description of neural networks. In particular, the method might apply to description of microcolumns, e.g., the fairly uniform, stereotyped, vertical column-like architecture believed to be the basic embryological/anatomical module and the fundamental processing unit of cortical arrangement (Mountcastle 1997; Jones, 2000). Minicolumns are characterized by modular connectivity with invariant properties that resembles fullerene structures. Indeed, minicolumns display a translational symmetry across their central axis and rotational symmetry, *i.e.*, displacement in different planes of section. Furthermore, microcolumns are equipped both with transitive symmetry involving geometric scaling of morphometric relations in different cortical areas, and with temporal symmetry involving morphometric relations during cortical maturation (Opris and Casanova, 2014).
Because architectonic relations among minicolumnar elements (e.g., pyramidal cells) are conserved under spatial and temporal variations (Casanova et al., 2011), it is possible to assess the tubular structure of a microcolumn in terms of a 3D cylinder with the typical regular structure of the Fullerene, e.g.,a mesh of variously fused exagons and pentagons. When the cylindrical structure is flattened into a 2D rectangular sheet, we achieve a lattice that allows us to investigate microcolums' microcuircuitry and its operations in a way different from the classical logic of ON/OFF neural firing. Indeed, a Fullerene-like 2D sheet is a structure where countless transformations, dictated by accurate rules and constraints, might occur. This structure can be compared to a barcode, or a biological matrix, in which every sequence of neuronal activation stands for a mental activity. We may also think to the bronze cup-shaped bells of a carillon, in which every sequence of diverse punchers gives rise to different melodies. In sum, we achieve a fullerene-like brain which activity is dictated by topological transformations taking place just on hexagons and pentagons. Such an approach not only allows us to build a model that faithfully simulates the biological brain activity, but also allows us to assess neural computations in terms of topological relationships and transformations among spiking neurons, instead of logic gates series or parallel circuits. We also discuss the advantages of such an approach in terms of network fastness, robustness and plasticity.

**MATERIALS AND METHODS**

In this section, we provide the procedure in order to build a fullerene-like microcolumn, in which neuronal firing and electric signal propagation are assessed in terms of pure topological neural network modifications, instead of canonical ON/OFF logic circuits. At first, we illustrate how to shape a fullerene-like lattice through graph theoretical methods. Then, we evaluate the (fully reversible) topological moves taking place on such a lattice. Finally, we assess the neural biological counterparts of fullerenic structures, assessing microcolumns and neural spikings in terms of movements on proper 2D graphs. Furthermore, we provide, in the section "RESULTS", a couple of simulations in order to test the feasibility of our fullerene-like microcolumnar framework.

**Fullerene networks come into play**. Fullerenes are 3-connected cubic planar graphs consisting of pentagons and hexagons (Schwerdtfeger et al., 2013). Geometrically, a fullerene is a closed trivalent polyhedral network in which atoms are arranged in 12 pentagonal and $\left(\frac{1}{2}n-10\right)$ hexagonal rings (Fowler et al., 1993; Fowler et al., 2001; Chuang et al., 2009). A regular 2D fullerene is a 3-regular cubic planar graph F that is generally describable as a mesh of (exactly) 12 variously fused pentagons, and surrounded by distorted graphenic fragments. A sample fullerene is shown in **Figure 1**.

Signal propagation can be investigated through graph theoretical methods. The basic assumption is that the evolution of microcircuits is ruled by the minimization of specific topological invariants (or topological potential) $\Xi$. Among the countless possibile choices (Todeschini and Consonni, 2000), we focus here on *distance-based* topological indices able to describe long-range interactions involving all pairs of nodes. The microcircuit is described as a graph $G_n$ with $n$ nodes in which the $d_{ij}$ integers stands for the number of graph edges connecting the two nodes $i$ and $j$ along the shortest path. In case of the distance $d_{ij}=k$, the node $j$ belongs to the $k$-th coordination shell of $i$, and vice-versa. By definition, $d_{ij}=d_{ji}$ and $d_{ii}=0$ for all nodes $i,j$.

If we term $M$ the length of the longest path of the graph - the integer $M$ stands for the graph *diameter* - and $b_{ik}$ the number of $k$-neighbors of $i$, the effects of the long-range connectivity on a single node is summarized by the topological invariant $w_i$:

$$w_i = \tfrac{1}{2} \Sigma_k k b_{ik} \qquad k=1,2,\ldots,M\text{-}1,M \qquad (1)$$

where $n=\Sigma_k b_{ik}+1$ and $b_{i1}=6$ for any 6-fold node $i$. The symbols $\underline{w}$ and $\overline{w}$ indicate the smallest and the largest $w_i$ values. Nodes with $w_i=\underline{w}$ (or $w_i = \overline{w}$) are the *minimal nodes* (or, conversely, the *maximal nodes*) in $G_n$. Integers $\{b_{ik}\}$ identify the Wiener-weights (WW) of the node $i$. The Wiener index $W$, that is the first topological descriptor applied in chemistry more than 60 years ago (Todeschini and Consonni, 2000), derives from the half-summation of $d_{ij}$ entries:

$$W(n) = \tfrac{1}{2} \Sigma_{ij} d_{ij} = \Sigma_i w_i \qquad i,j=1,2,\ldots,n\text{-}1,n \qquad (2)$$

The first topological modeling rule is the following: higher reactivity is assigned to a node with maximal $w_i$.

$W$ provides the topological measure of the overall compactness of the system and is the first good candidate for the topological potential $\Xi^W=W(G_n)$. This approach is grounded on relevant heuristic evidences, such as the fact that, among the 1812 non-isomorphic $C_{60}$ fullerene isomers, just the physically stable isomer with icosahedral symmetry $C_{60}$-$I_h$ and isolated pentagons corresponds to the cage with the minimum value W=8340.
The following invariant :

$$\rho = W/n\underline{w} \qquad (3a)$$

measures instead the overall ability of the graph to evolve as a compact topological structure around its minimal nodes, which are considered the most efficient nodes in $G_n$. From formula (3a), this the important inequality follows:

$$\rho \geq 1 \qquad (3b)$$

For this reason $\rho$ is called a *topological efficiency* index (Ori et al., 2009), that has been recently used in order to rank the *topological sphericity* of fullerene graphs (Graovac et al. 2014). Many chemical structures have $\rho=1$. Infinite cubic lattices or infinite graphene layers, together with icosahedral $C_{60}$ molecules, are examples of perfectly spherical (topologically speaking) structures. It follows the second topological modeling rule: higher stability is assigned to structures with maximal sphericity, e.g. minimal $\rho$. The two above mentioned modeling rules hold when similar topological structures are compared. The lattice descriptor $\rho$ will also stand in the present study for the second form of the topological potential $\Xi^\rho = \rho\,(G_n)$ able to rule signal propagations in the mesh.

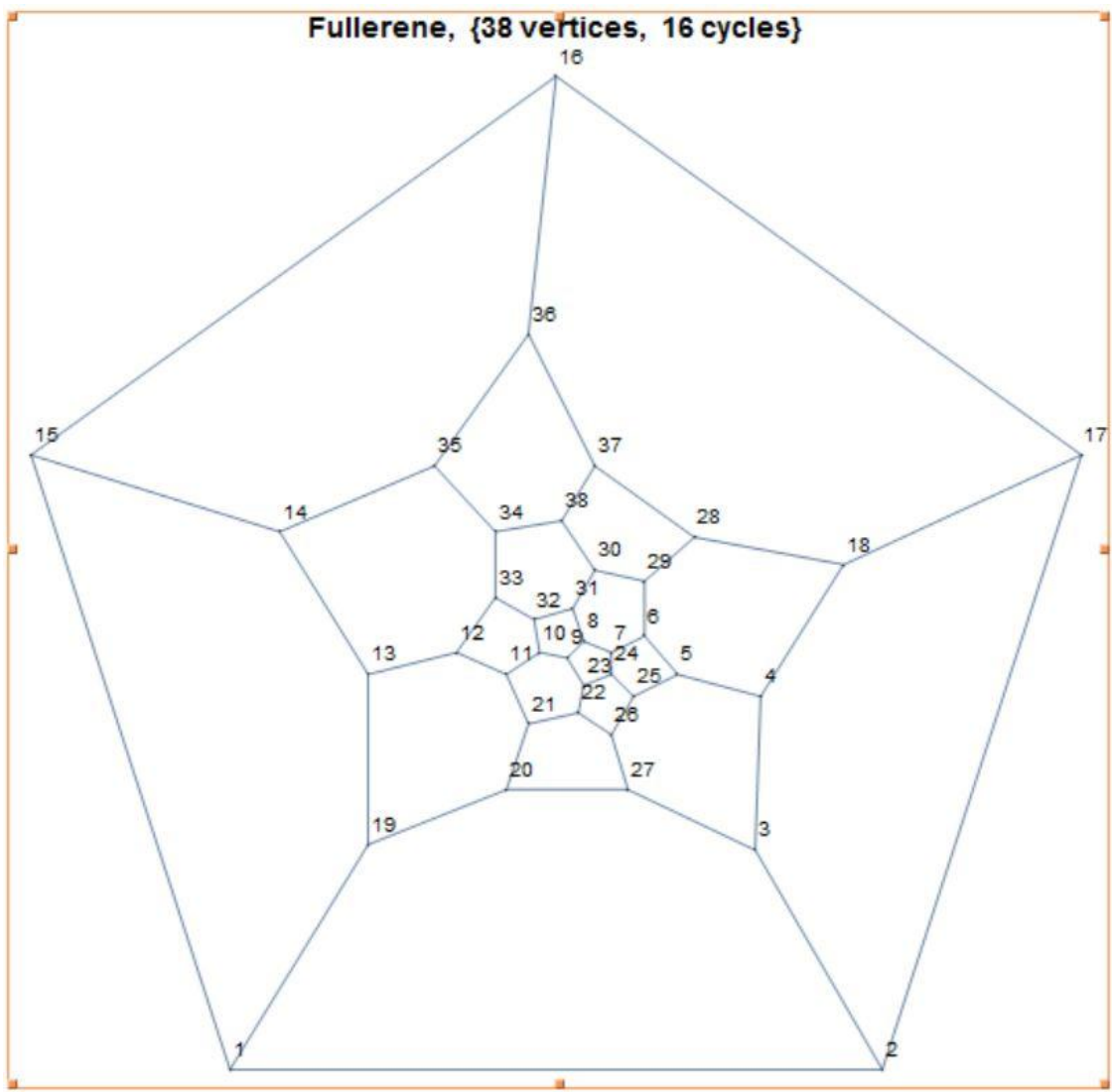

**Figure 1**. Sample fullerene with 38 vertices and 16 cyles. See text for further details.

**Stone-Wales transformations in neural networks.** As stated above, in this paper neuronal firing and electric signal propagation are investigated in terms of pure topological neural network modifications. This assumption implies that the overall number of neurons $n$ (graph vertices) and connections $B$ among neurons (graph edges or bonds) are *fully preserved*, excluding that neurons are created or destroyed along the process. Moreover, local deformations should not affect the topological structure of the neural lattice's surrounding regions. Because such kinds of topological transformations are totally reversible, they provide a natural topological mechanism for signal annihilation or suppression. Topological permutations occurring on the hexagonal or pentagonal surface of a fullerene-like lattice can be assessed and calculated. Indeed, methods based on vertex insertions or deletions and isomerisations on lattices have proven to be very useful, because they allow the formal derivation of one fullerene graph from another (Ori et al., 2014). In particular, the so-called Stone-Wales (SW) transformations produce a rotated figure, without changes in the external connectivity of the fullerene and in the molecular topology outside the represented region of the lattice surface (Babic et al., 1995). In sum, the SW mechanism makes a dipole dislocation extending in the fullerenic mesh as a wave. Stone-Wales (SW) transformations might play an important role in modifying not just the topology of carbon-based materials, like carbon nanotubes, fullerenes and other two-dimensional materials (Ma et al., 2009), but also the properties of neuronal circuits.

Here we show how to operationalize the procedure. The general local Stone-Wales transformation $SW_{q/r}$ (**Figure 2**) is associated with the isomeric transformation that modifies the internal connectivity of four adjacent rings with $p, q, r, s$ edges, producing four new rings with $p–1, q+1, r–1, s+1$ edges, without changes in the surrounding regions of the lattice. $SW_{p/r}$ reversibly rotates the central bond bridging $p$ and $r$ rings, leaving unchanged both the total number of nodes $k=p+q+r+s-8$ and the total number of edges k+3. **Figures 2B** and **2C** illustrate the two most studied fullerenic forms where a rotation of a pentagon-hexagon double pair occurs. SW rotations are known to be an important tool in order to generate fullerene isomers with different symmetries. In the crucial case of the $C_{60}$ fullerene, the *pyracylene* rearrangements $SW_{5/6}$ group the 1812 isomers in 13 inequivalent sets (**Figure 2C**). In the largest of these sets, 1709 isomers are connected to the buckminsterfullerene cage through one or more SW flips, leaving 31 isomers unconnected to any of these sets. In an early work on this topic (Ori et al, 2009), this limitation has been solved by using extended generalized Stone-Wales transformations, in order to produce, starting from just one isomer, the whole $C_{60}$ isomeric space.

In a recent study (Ori et al, 2011), a convenient graphical tool has been developed, in order to modify fullerene-like networks displayed as 2D graphs made by $n$ vertices (*carbon atoms or neurons*), through the isomeric insertion of different $r$-rings with $r>2$. The case $r=5,6$ stands for the regular fullerene, while $r=6$ for the pristine graphene. The efficiency of the graphical method relies on the dual-representation of the graph, that can be achieved by transforming the rings in nodes and their edges in bonds. **Figures 2D** and **2E** provide some examples of the interplay between direct and dual representations of fullerene-like graphs. The straightforwardness in generating SW rotations is self-evident. The efficiency of the graphical tool allows, as a consequence of iterated SW rotations, the identification of a peculiar topological mechanism called the Stone-Wales wave (SWw) (Ori et al, 2011), consisting in the linear migration of the pentagon-heptagon 5|7 pair in the fullerene-like planar network. The first rotation $SW_{6/6}$ of the connection (arrowed) shared by the two neurons creates two 5|7 pairs (**Figure 2D**). The second operator $SW_{6/7}$, by inserting the 6|6 couple of shaded hexagons between the two original 5|7 pairs, turns the bond between the heptagon and the shaded hexagon (**Figure 2E**). This mechanism produces the overall effect of initiating the wave propagation along the dotted direction, because iterated $SW_{6/7}$ rotations will successively propagate the topological defect 5|7. Concerning the mathematical apparatus related to symmetries and transformations on nanostructures, see also Koorepazan-Moftakhar et al (2015).

**Building a fullerene-like microcolumn.** Microcolumnar circuitry can be described in terms of fullerene 2D lattices, equipped with the same asymptotic topological behaviour. A cortical microcolumn comprises about 80–120 neurons, except in the primary visual cortex where the number doubles. About 80% of such cells are pyramidal neurons. This basic unit circuit forms a series of repeated items across the horizontal extent of the cortex, independent of cortical areal specializations (Jones, 2000): in humans, the transverse diameter of each of the $2x10^7$-$2x10^8$ microcolumns is about 28–50 μm (Sporns et al., 2005; Johansson and Lansner, 2007). Therefore, we are allowed to build a biologically-plausible, fullerene-like lattice describing a microcolumn. **Figures 3A** and **3B** display further details about the building steps from tubular cortical microcolumns to fullerene-like two-dimensional sheets. In sum, we achieve a 30 microns-width lattice equipped with exagonal (or pentagonal) tiling and about 80 nodes (or vertices). Such nodes, standing for the about eighty microcortical pyramidal neurons, are linked by edges (or bonds), e.g., dendritic or axonal connections. Approaching SWw propagation in neuron circuits gives rise to phenomena, including the sensitivity to the size of the circuit with an inherent anisotropy, that may be investigated by a topological perspective, namely by considering the topological potential of neural networks. Therefore, SW transformations might have a counterpart in the real biological activity of microcolumnar microcircuitry. In the nervous case, instead of SW flips on a fullerene surface, we could take into account the changes in firing activation of the about 80 neurons embedded in the nodes. In the case of microcolumns, *permutation* means which pyramidal neurons are sequentially activated and which are deactivated. In other words, SW permutations may be regarded as a set of simple pyramidal neurons which fire simultaneously (**Figure** 3C). Every group of neural "permutations" might stand for a sort of barcodes, each one corresponding to a peculiar mental activity.

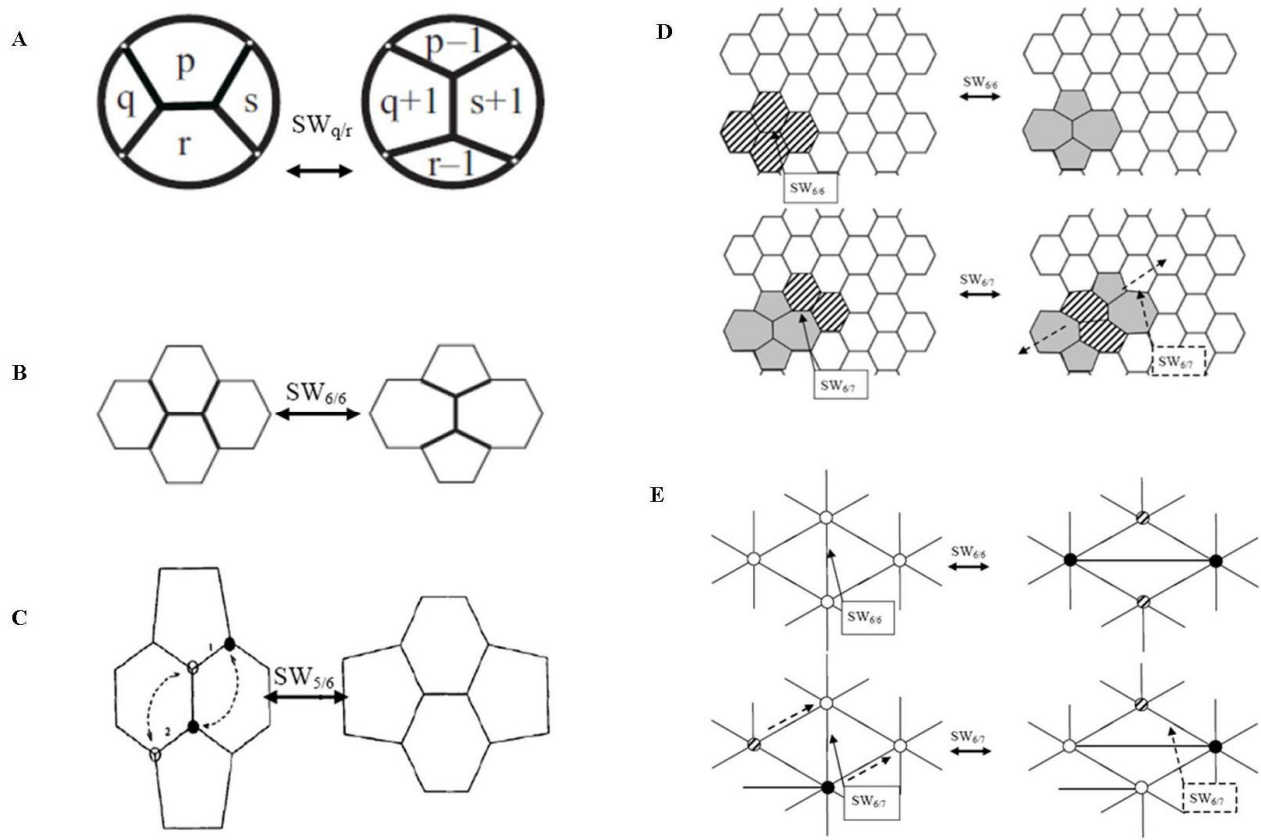

**Figures 2A-2C**. Examples of Stone-Wales transformation $SW_{q/r}$. **Figure 2A**: the general $SW_{q/r}$ rotation reversibly transforms four proximal polygons with p, q, r, s edges in four new rings with p–1, q+1, r–1, s+1 nodes. **Figure 2B**: in the graphene p=q=r=s=6, standing for the $SW_{6/6}$ Stone-Thrower-Wales rotation normally encountered in graphene and carbon nanotubes, $SW_{6/6}$ flips four hexagons in a 5|7 double pair. **Figure 2C**: the so-called "pyracylene rearrangement $SW_{5/6}$" displays a $SW_{5/6}$ rotation on the fullerene surface.
**Figures 2D-E** summarize the basic topological operations for generating and propagating Stone-Wales waves in the lattice. **Figure 2D**: in the direct graph, $SW_{6/6}$ originates two 5|7 pairs (gray), then $SW_{6/7}$ splits them, by swapping one of the 5|7 two pairs with a pair of hexagons (shaded); dotted $SW_{6/7}$ propagates the SW wave in the dashed direction. **Figure 2E** displays the same transformations in the dual plane. Hexagons, pentagons, heptagons are represented by white, shaded, black circles respectively. See text for further details.

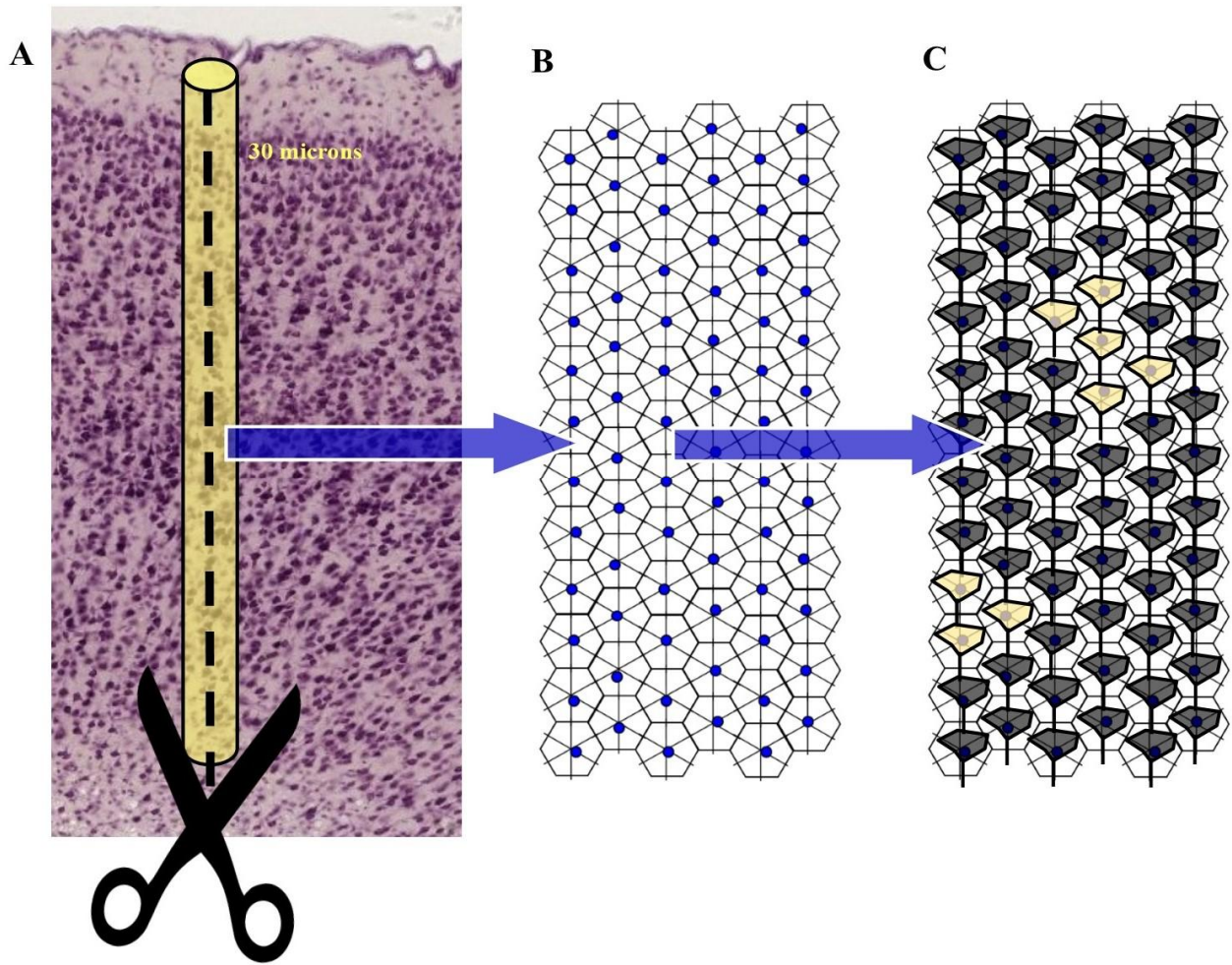

**Figure 3**. General description of a fullerene-like cortical microcolumn. In **Figure 3A**, the anatomical/functional basic structure of the brain, e.g., the tiny cortical microcolumn, is shaped in guise of a three-dimensional tubular armchair. If we cut the column through an edge, we achieve a flattened two-dimensional fullerene-like grid with exagonal tiling (**Figure 3B**). In the hypothetical case of **Figure 3B**, provided just as an example, the molecular graph stands for a 2D zig-zag lattice, e.g., a flat, open visualization, made of 72 nodes with 6-bonds each. **Figure 3C** depicts a few possible topological transformations that might take place on this flattened microcolumn. Every node is filled with a pyramidal neuron which can be activated (yellow shapes) or deactivated (black shapes). Every set of activated/deactivated neurons gives rise to a different microcolumnar barcode, each one standing for a single mental activity among the countless possible.

**RESULTS**

Here we provide some simulations that show how, starting from a fullerene structure, it is possible to reversibly generate many alternative isomers with a lower structural symmetry, just by twisting two hexagons around a central bond (**Figures 4** and **5A**). The possible combinations of permutations (and therefore of simultaneously activated neurons in a fullerene-like microcolumn) are countless: in a lattice of 80 nodes and 2 activities (ON/OFF), we achieve the noteworthy number of $2^{80}$ possible permutations. Many operations, not shown in Figures, can be performed: edges sharing, interchanging the positions of two hexagons, rotation of a *single* edge, and so on. Because the switches are fully reversible, the edges can be rearranged in countless ways.

It might be objected that we do not know, due to our current lack of knowledge, whether the microcortical assembly of pyramidal neurons exhibits the required fullerene-like stereotyped conformation. For example, in our simulation in **Figure 5A**, the lattice exagons appear to be "stretched". The Borsuk-Ulam theorem (BUT) from algebraic topology, ideally suited for many applications (Matoušek, 2003; Blagojevic and Ziegler, 2016), gives us a clue to solve this problem. The original formulation of BUT describes antipodal points with matching description on spatial manifolds in every dimension, provided the n-sphere is a convex structure with positive curvature (i.e, a ball). However, BUT can be generalized to symmetries occurring either on flat manifolds, or on Riemannian hyperbolic manifolds of constant sectional curvature -1 and concave shape (i.e, a saddle) (Mitroi-Symeonidis, 2015; Tozzi and Peters, 2016). In other words, whether the system components are equipped with concave, convex or flat conformation, it does not matter: we may always find the points with matching description predicted by BUT (Tozzi, 2016). This means that, even if the pyramidal neurons are not arranged into the microcolumn in guise of a perfect fullerene-like structure, we can always describe their activations in terms of transformations on two-dimensional zig-zag lattices (**Figure 5B**).

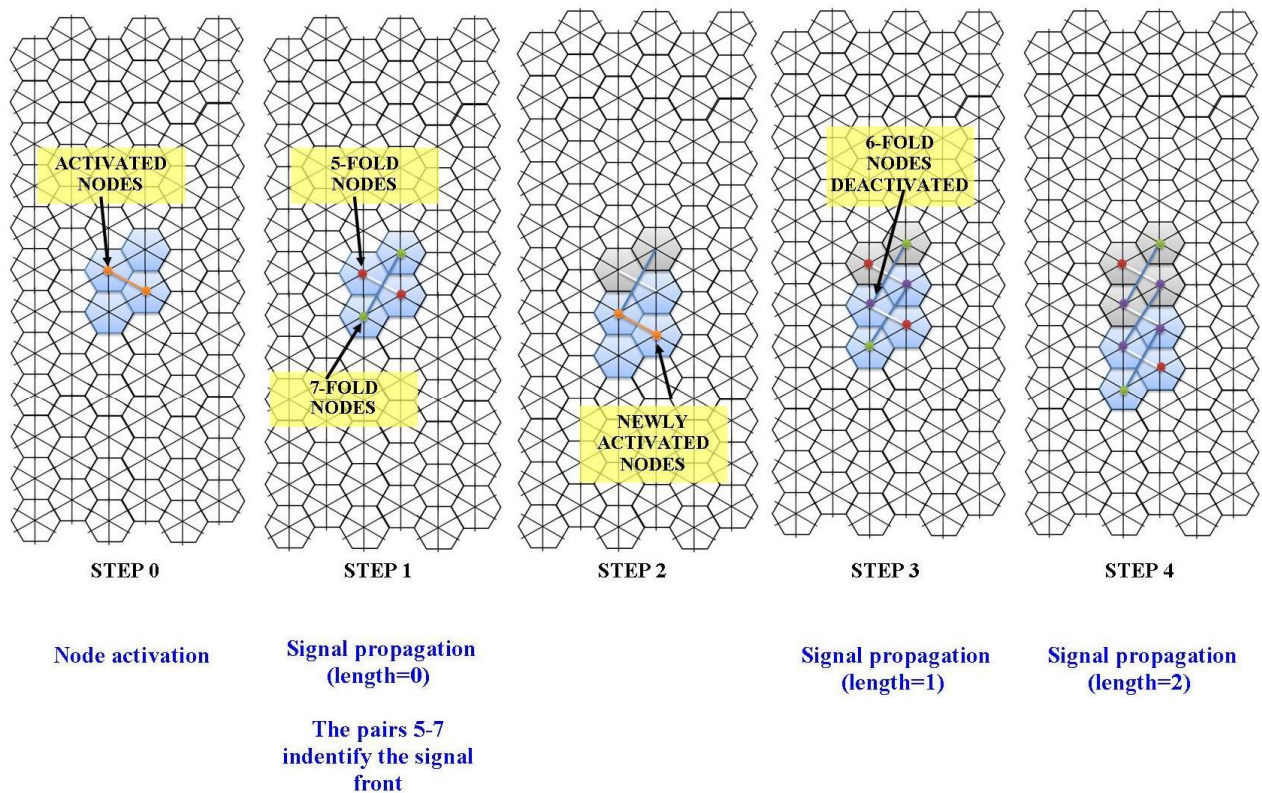

**Figure 4**. Simulation of a few possible transformations taking place on a flat exagonal lattice equipped with 72 nodes, 6-bonds each. The lattice stands for the cortical microcolumn containing 72 pyramidal neurons described in **Figure 3C**. Starting from thetotally inactive state of 6-fold nodes on 6x12 lattice (displayed in **Figure 3B**), some nodes gets activated (**STEP 0**). We may select whatever pair in the mesh. To make an example, after a visual input from the external environment, a few microcolumnar pyramidal neurons start to fire. **STEP1** displays two 5|7 pairs, activated by just rotating the orange bond of **STEP 0**. Signal propagation gives rise to a sequential activation of novel front nodes (**STEP2**), so that the activation wave propagates steps further (**STEPS 3** and **4**). Note that the purple balls, which identify hexagonal nodes deactivated after the passage of the wave, are ready for a possible reversible annihilation of the topological signal.

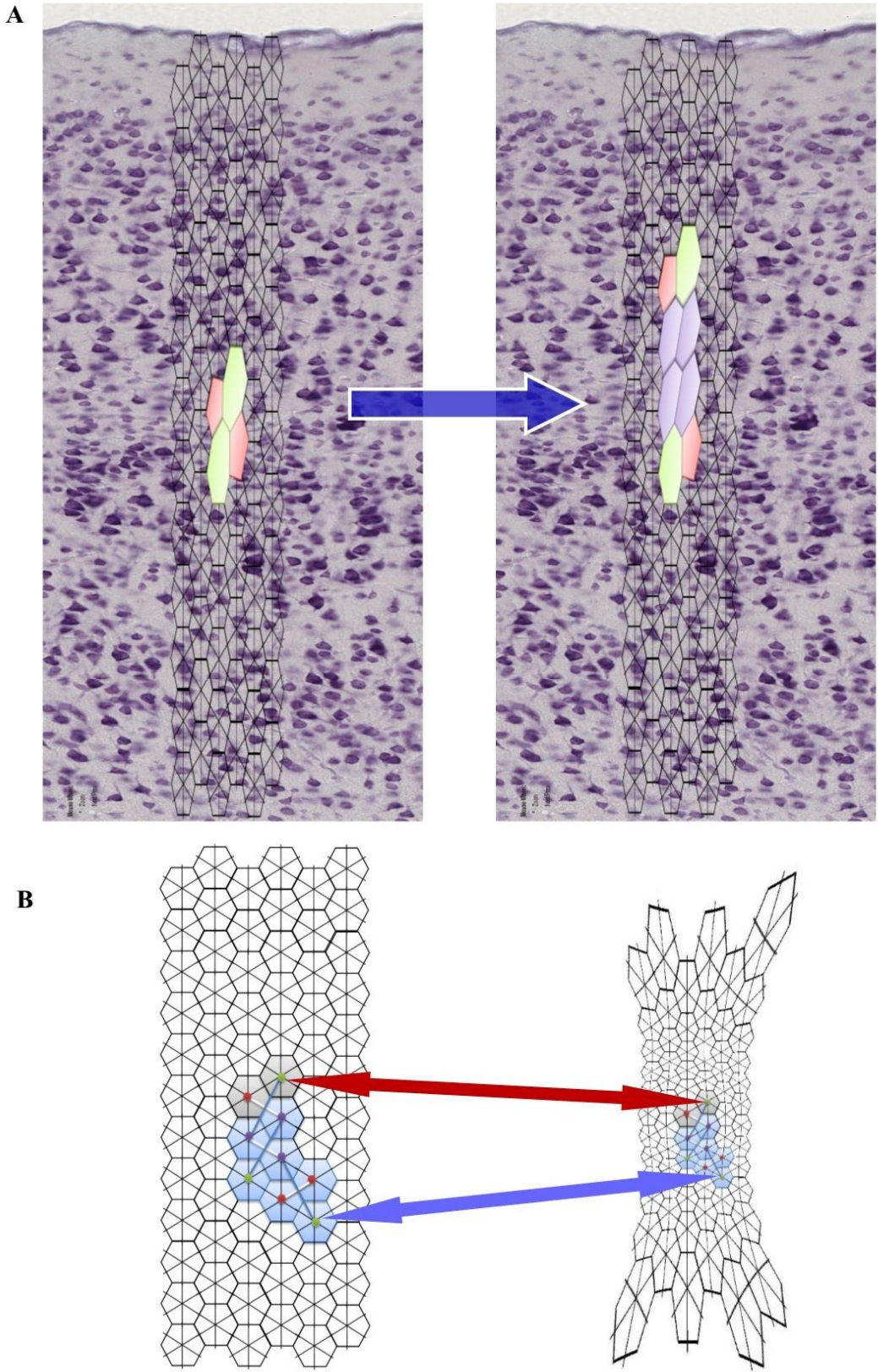

**Figure 5A**. Topological simulation of propagating signals on a fullerene-like grid embedded in situ into a cortical micrococolumn. The stained exagons, depicting SW transformations in lattice's nodes, stand in this case for activated pyramidal neurons. See main text for further details. **Figure 5B**: every kind of transformation in a fullerene molecule leaves the relationships among the single elements unchanged, do to the Borsuk-Ulam theorem's dictates. The Figure shows how a deformation of the original fullerene-like lattice leaves the location of firing neurons unchanged.

**DISCUSSION**

We achieved a 2D fullerene-like lattice reproducing microcolumn's microcircuitry. The question now is: given the countless nervous computational models (Van den Heuvel and Sporns, 2011; Friston, 2010; Sengupta et al., 2016), what does a fullerene-like brain theory bring into play, in the evaluation of brain activity? Although our lattice looks like a classical neural network at very first sight, nevertheless it is faster, less energy-consuming, more plastic and stabler. It is also easy to parametrize, because information can be extracted from just a few parametrizing factors. e.g., the simple topological operations taking place on our network. While other approaches emphasize macro-phenotype dynamics, our framework is grounded on dynamics occurring at nervous micro-levels. We assume that the simultaneous firing of different neurons and specific activation sequences might give rise to different topological conformations, each one corresponding to a mental activity (perception, emotion, mind-wandering, calculation, and so on). The brain functional activity is not based on logic nodes as it occurs in conventional networks, but on topological transformations, e.g., functional changes in the position of firing neurons on a barcode.

Our model explains countless mental operations starting just from a relatively simple, stereotyped, highly preserved biological structure, such as the microcolumn. Fullerene-like models, involving any pair of connected nodes (Maruyama and Yamaguchi, 1998), unveil a very rich phenomenology, due to the almost infinite SW rotations that may take place on the corresponding lattice. This makes possible a limitless set of moves, merely dictated by applying SW rotations on its bonds. Ubounded sets correspond to the infinite series of mental operations taking place in our brain. During the syntesis and growth process of a $C^n$ fullerene, the isomers with isolated pentagons often udergo SW interconversion. If we use a given $C_n$ isomer as a seed, we may perform a series of moves that give us the capability to build complete sets of fullerene isomers, by simple binding topological briges (Babic et al., 1994). Therefore, both isomerization maps and complete isomer space's description can be achieved from a single known conformation. In the same guise, the primal embryonal microcolumn is able to give rise to countless combinations in microcircuitry in the mature individual. Indeed, starting from the single vertical template of about 80 pyramidal neurons that colonizes the whole cortex during embryonal life, countless barcodes might be build with time passing. In touch with catalysts that promote fullerene transformations able to stabilize its transitions states, we hypothesize that the habituation to repeated stimuli during the first years of life could give rise to more stable and permanent transformations of the microcolumns barcode.

Fullerene-like networks are fast, because they use simple transformations. Starting from each randomly built string of $C_n$ fullerene, one may directly generate a certain number of new isomers using just SW flips, instead of immediately searchingh for a new one. It improves computational speed, because the number of steps required by every operation is reduced. A simple increase in the order of the SW transformations allows us to connect more microcolumns: this means that dual structures, given by the combination of many microcolumns, can be easily built by joining the reciprocal space. In the same way as the tuning of the nanoparticle/substrate interaction provides unique ways of controlling the nanotube synthesis, fullerene-like, microcolumnar neural networks might provide the fine-grain functional barcode's modularity required by different brain functions (Gomez-Ballesteros et al., 2015). It means that modifications in reciprocal connectivity among adjacent microcolumns allow a larger repertoire of barcode configurations and, therefore, of mental operations.

Fullerene-like networks take time to become efficient, because SW transformations are less favoured by an energetic point of view. Therefore, the building of novel barcodes is a slow, time-consuming process. By a biological point of view, this is in touch with the slow development of mental operations' competence in young Primates (Skeide and Friederici, 2016). Vice versa, a fullerene-like biological neural network gives to adult Primates unvaluable advantages in terms of energy sparing. Incorporation of pentagons, occurring at an early stage of nucleation pathway for for single-wall nanotubes, leads to a very efficient system, because the number of dangling bonds is reduced and changes towards more entropic curvatures are achieved (Fan et al., 2003). To make an example, fullerenes produced by the overlap of 12 nanocones which no direct pentagon fusion display very high thermodynamic stability, steady topological configurations and minimized graph invariants (Vukicevic et al., 2011; Ori et al., 2014). In mental terms, a fullerene-like microcolumnar structure is associated with minimal wiring costs and fast synchronization/information transfer (Stam and Reijneveld, 2007). In touch with neural networks derived from the principle of minimum frustration for protein folding (Tozzi at al., 2016), the fullerene-like approach points towards a perception/decision apparatus constrained towards very low energetic levels, but just in long timescales, in order that predictions signals are conveyed by the sole long-standing past experiences.

In sum, we make available a novel approach that allows us to evaluate brain activity in terms of topological and graph theoretical properties of fullerenes. This is a very fertile field of research: current efforts focus on the unsolved mathematical problems concerning fullerenes, e.g., how to generate all possible non-isomorphic graphs for a fixed vertex count, or to calculate the the number of distinct Hamiltonian cycles. Researchers are developing 2D graphs and 3D structures for many different fullerenes, ranging from $N= 20$ to 20,000 vertices, in order to evaluate various different graph-theoretical algorithms (Schwerdtfeger et al., 2015). Theorems and algorithms which allow a fast computation of topological indices for complex graphs, starting from their structural building elements, are effortlessy produced (Koorepazan-Moftakhar et al, 2015). Such powerful tools could be, in a near future, applied to the assessment of neural networks, in order to achieve a double task: improving our understanding of biological brain function in situ, and building more powerful artificial machines able to simulate cortical activities. Our framework also provides a link with human neurologic and psychiatric diseases. Experimental data and molecular dynamic simulations suggest that defects during the nucleation and growth of graphene alter the physical/chemical properties of carbon nanostructures (Hashimoto et al., 2004), strongly deteriorating their functional state. An increase in energetic constraints, as it occurs, for example, during ageing or central nervous system's diseases, makes it easy to generate defects in the fullerene-like lattice, with subsequent decrease in system's free-energy and functional integrity. Our theoretical framework, cast in a biologically informed fashion, has the potential to be operationalized and assessed empirically. Indeed, the presence of a microcolumnar barcode predicted by a fullerene-like brain will be easily testable, once more powerful high density neurotechniques will be available, capable of capturing the simultaneous activity of large populations of microcolumnar pyramidal neurons, (Koster et al., 2014). To make an example, if we could be able to evaluate the specific microcolumnar pyramidal neurons which fire during every mental activity (e.g., visual tasks, emotions, and so on) in order to achieve a fullerene-like structure filled with the corresponding activated nodes, we could attain a series of different grids or matrices, each one standing for a mental function. We suggest to analyize primate temporal cortex or other associations cortices, instead of the widely studied rodents' barrel cortex or the primates' visual cortex, because the former display a notable sterotypy (Jones, 2000), which may prove to be a better model.